\documentclass[preprint]{ptephy_v1}

\preprintnumber{XXXX-XXXX} 

\begin{document}

\title{Inverse Mass Hierarchy of Light Scalar Mesons \\ 
Driven by Anomaly-Induced Flavor Breaking}


\author{Yoshiki Kuroda}
\author[1]{Masayasu Harada}
\affil{Department of Physics,  Nagoya University, Nagoya, 464-8602, Japan}

\author{Shinya Matsuzaki}
\affil{Center for Theoretical Physics and College of Physics, Jilin University, Changchun, 130012, China}

\author{Daisuke Jido}
\affil{Department of Physics, Tokyo Institute of Technology, Meguro, Tokyo 152-8551, Japan
\email{kuroda@hken.phys.nagoya-u.ac.jp}
\email{harada.masayasu@nagoya-u.jp}
\email{synya@jlu.edu.cn}
\email{jido@phys.titech.ac.jp}
}



\begin{abstract}%
We propose a novel mechanism to reproduce 
the observed mass hierarchy for the scalar mesons 
lighter than 1 GeV (called the inverse hierarchy) regarding them 
as mesons made of a quark and an anti-quark ($q\bar{q}$ mesons).
The source is provided by the SU(3)-flavor symmetry-breaking induced by U(1) axial anomaly.
In particular, 
the anomaly term including the explicit-chiral symmetry-breaking plays
a significant role for the light scalar meson spectrum.
To be concrete, we construct
a linear sigma model for the scalar mesons of $q\bar{q}$ type
together with their pseudoscalar chiral partners, 
including 
an anomaly-induced explicit-chiral symmetry-breaking term. 
We find that, due to the proposed mechanism, the inverse hierarchy, 
i.e., $m\left[ a_0 (980) \right] \simeq m\left[ f_0 (980) \right] > m \left[ K_0^\ast (700) \right] > m \left[ f_0(500) \right]$ 
is indeed realized. 
Consequently, the quark content of $f_0 (500)$ is dominated by the 
isoscalar $\bar uu+ \bar dd$ component, 
and $f_0 (980)$ is by the strange quark bilinear's, $s\bar{s}$.
\end{abstract}

\subjectindex{D32}


\maketitle

\section{Introduction}

The vacuum structure in QCD is supposed to be 
governed by the nonpreturbative quark condensate, 
which spontaneously breaks the (approximate) chiral symmetry. 
It gives rise to the associated (pseudo) Nambu-Goldstone bosons, 
arising as pseudoscalar mesons in the meson spectra. 
In that sense, the dynamics among those pseudoscalar mesons is 
governed by the spontaneously broken chiral symmetry, 
which is well described in the framework of 
the chiral perturbation theory~\cite{Weinberg:1978kz,Gasser:1983yg,Gasser:1984gg}. 
  
In addition to those pseudoscalar mesons,   
scalar mesons made of a quark and an anti-quark, 
so-called $q\bar{q}$ scalar mesons, 
are also expected to exist as their chiral partners, 
which should include the fluctuation mode of the flavor-singlet chiral condensate, 
regarded as the signal particle for the chiral symmetry breaking 
(conventionally denoted as sigma meson), 
similarly to the Higgs boson for the electroweak symmetry breaking. 
Thus, it is important to specify the $q\bar{q}$ scalar mesons 
in the scalar meson spectra, 
which would provide a clue to deeply understand the low-energy QCD, 
as well as the vacuum structure in terms of the chiral symmetry breaking, 
and also some hints for the origin of hadron masses. 

Possible candidates for the $q\bar{q}$ scalars 
would involve the low-lying scalar mesons with masses below 1 GeV, 
which are $a_0 (980)$, $f_0 (980)$, $K_0^\ast (700)$ and $f_0(500)$. 
It is interesting to note that
the masses of these scalar mesons satisfy the so-called ``inverse hierarchy'':
\begin{align}
\label{inverse_hierarchy}
m\left[ a_0 (980) \right] \simeq m\left[ f_0 (980) \right] > m \left[ K_0^\ast (700) \right] > m \left[ f_0(500) \right] \,. 
\end{align}
This implies  
some significant effect of the SU(3) flavor (nonet) breaking due to 
the mass difference for the up, down and strange quarks, 
in which the strange quark mass would presumably give 
a dominant flavor symmetry breaking. 
However, the observed hierarchy in Eq.~(\ref{inverse_hierarchy})
is actually contradicted to a naive observation
obtained just by counting the number of (valence) strange quarks:
$
m\left[ \sigma_s \right] > m\left[K_0^\ast\right] > m\left[\sigma_n\right] \simeq m\left[a_0\right] 
$,
where $\sigma_n$ is composed of up and/or down quarks, 
and $\sigma_s$ of strange quarks.
Then, it might be hard to simply identify the light scalar mesons 
as the $q\bar{q}$ states, instead, they might rather be 
four-quark ($qq\bar{q}\bar{q}$) mesons or mixture of $q\bar{q}$ and $qq\bar{q}\bar{q}$. 
In fact, this kind of argument has urged people to work on some complicated scalar 
meson puzzle, which has currently been a major approach in the ballpark of this sort    
(see e.g., Refs.~\cite{Pennington:2007yt,Fariborz:2009cq,Klempt:2007cp,Pelaez:2015qba,
Achasov:2017ozk,Tanabashi:2018oca,Jaffe:1977,Black:1998wt,
Chen:2006zh,Zhang:2006xp,
Prelovsek:2008rf,Prelovsek:2010kg,Wakayama:2014gpa,
Fariborz:2005gm,Mukherjee:2012xn,Albaladejo:2012te,
Fariborz:2011in,Fariborz:2015bff,Pelaez:2016klv,Pelaez:2017sit} 
and references therein).   

Still, some analyses have been attempted without getting into such a complexity: 
Reference~\cite{Naito:2002} studied the masses of the scalar mesons in
an extended Nambu-Jona-Lasino model including 
a six-quark interaction term induced by the U(1) axial anomaly 
a l\'a Kobayashi-Maskawa-t'Hooft (KMT)\cite{Kobayashi:1970,Kobayashi:1971,
tHooft1976,tHooft:1976L,Rosenzweig:1979ay,DiVecchia:1980yfw,Witten:1980sp}. 
It was shown that an $I=0$ (isosinglet) $q\bar{q}$ scalar 
meson, which is a singlet under the SU(3) flavor symmetry, 
becomes lighter than the other scalar mesons belonging to an octet
due to the anomaly-induced interactions, 
which is consistent with the hierarchy between $f_0(500)$ 
and the others in Eq.~(\ref{inverse_hierarchy}).
In the model the SU(3) flavor violation is turned on just by 
the current quark mass term to be indirectly transferred 
into the meson spectra, 
through the dynamical quark masses and the quark condensates 
appearing in the quark propagator and the six-quark KMT interaction.
In particular, thanks to the flavor singlet nature of the KMT interaction,
the vertices contain the strange quark condensate, $\langle \bar ss \rangle$, for the $a_{0}$ meson
and the up or down quark condensates, $\langle \bar nn \rangle$,
for the $K_{0}^{*}$ meson. This contribution is against the simple strange 
quark number counting and works for resolving the inverse hierarchy. 
However, the $a_0$ meson was predicted to be still lighter than 
the $K_0^\ast$ meson, in disagreement with the observation, 
due to insufficient flavor symmetry breaking.
Recent works~\cite{Osipov:2012kk,Osipov:2013fka} showed that 
an NJL-like model extended by including a large number of explicit-flavor violating terms 
with the current quark masses can reproduce the 
"inverse hierarchy" for the masses of the scalar mesons.

The work in Ref.~\cite{Ishida:1999},
as an example which reproduces the inverse hierarchy,
used a linear sigma model (LSM) including the
U(1)  axial anomaly-induced term of KMT-like determinant type, 
and showed that the ``inverse hierarchy'' is realized. 
In the model the flavor violation structure is 
similar to the model in Ref.~\cite{Naito:2002}.
As a result, this model predicts a larger $f_K/f_\pi$ 
(the kaon decay constant over the pion decay constant) than the experimental value, 
which seems unsatisfactory. 
Note also that the large $f_K/f_\pi$ can be rephrased as 
a significantly larger size of the deviation for quark condensates from unity,  
$\left<\bar{s}s\right> / \left<\bar{n}n\right> \gg 1$,  
which has currently been disfavored by the lattice simulation result~\cite{McNeile:2013}. 
Thus, the key mechanism to realize the ``inverse hierarchy'' for masses is not yet 
clarified with regarding the light scalar mesons as $q\bar{q}$ mesons, to our best knowledge.

\begin{figure}[t]
\begin{center}
\includegraphics[bb = 0 0 360 214, width=7.5cm]{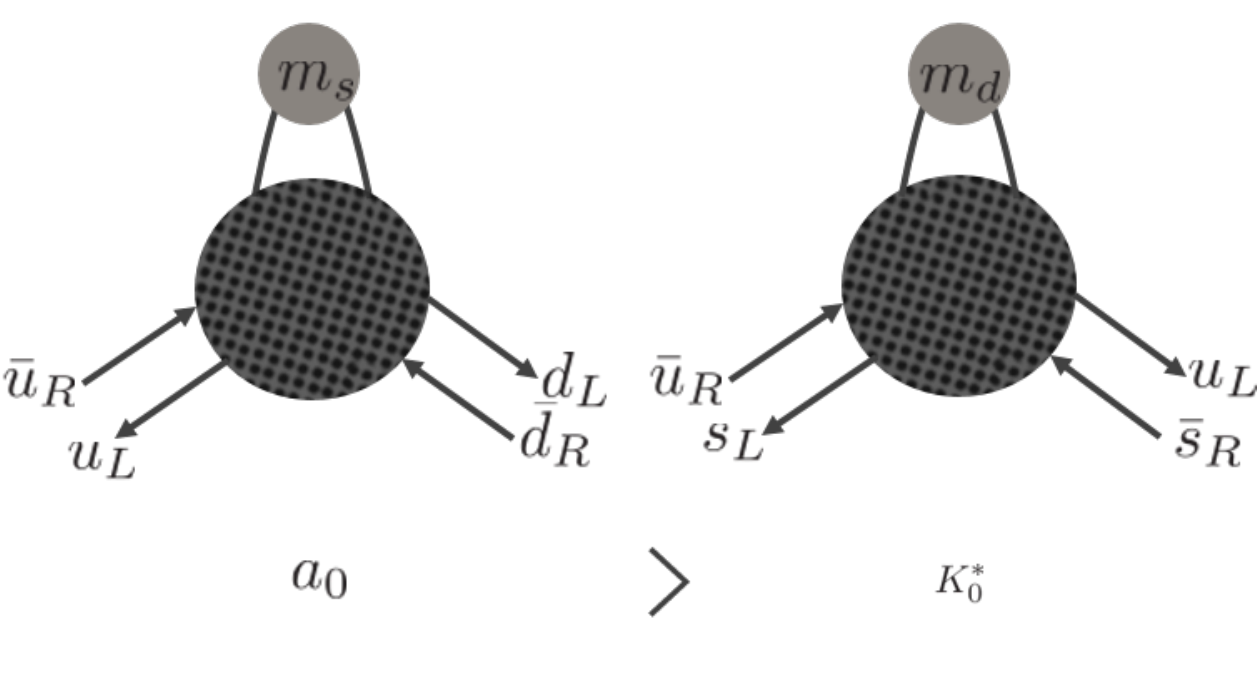}
\caption{
Schematic graphs illustrating contributions of the current quark masses 
($m_d, m_s$) 
to the masses of $a_0 (980)$ and $K_0^\ast (700)$ 
through the U(1)$_{A}$ 
anomaly-induced six-quark interaction of determinant type.
The blobs stand for the interaction vertex among six quarks 
with two of six quark lines being contracted and replaced by the current quark masses.   
It is clearly seen that due to the interaction of anomaly form,  
only the $a_0 (980)$ meson receives the mass correction proportional to 
the strange quark mass $m_s$, hence necessarily would get heavier 
than $K_0^\ast (700)$ by this mechanism.}
\label{fig:SU3-mechanism}
\end{center}
\end{figure}

In this paper, we propose a novel mechanism to realize 
the ``inverse hierarchy'' of the scalar meson masses.
The key ingredient is the explicit SU(3) flavor breaking 
due to the quark masses arising along with the U(1) axial anomaly,
which is counter to the quark number counting.
Figure~\ref{fig:SU3-mechanism} illustrates 
how the explicit-flavor breaking-anomaly interaction potentially 
makes $a_0 (980)$ heavier than $K_0^\ast (700)$ just by feeding the strange quark mass
in the same way as in the KMT interaction, 
in which the proposed interaction contributes to the $a_{0}$ 
meson with the strange quark mass due to the flavor singlet nature of the anomaly. 

This paper is organized as follows: In section~\ref{ELSM}, 
we will work on the LSM used in Ref.~\cite{Ishida:1999}, 
and extend it by introducing a new explicit SU(3)-flavor symmetry-breaking 
term reflecting the U(1) axial anomaly contributions involving the explicit-flavor breaking effect. 
In section~\ref{sec:massesl}, we show that the inverse hierarchy is indeed achieved, 
with keeping $f_K/f_\pi$ set to the experimental value, 
due to the presence of the anomaly-induced explicit flavor-breaking term. 
This manifests 
that the LSM with the explicit-flavor breaking-anomaly term 
improves the earlier work in Ref.~\cite{Ishida:1999}. 
We finally give a summary in section~\ref{sec:discussion}.
Formulas for the masses are summarized in  Appendix~\ref{app:masses}.

\section{Linear Sigma Model Description}
\label{ELSM}
We begin by introducing an 
LSM for the light scalar and pseudoscalar mesons 
made from the $q\bar{q}$ state. 
The meson fields are introduced so as to be embedded in 
a $3 \times 3 $ matrix field $M$, which transforms under 
$\mbox{SU}(3)_L \times \mbox{SU}(3)_R \times \mbox{U(1)}_A$ symmetry as
\begin{equation} 
M \to g_A \, g_L \, M \, g_R^\dag \ ,  
\end{equation}
where $g_{L,R} \in \mbox{SU}(3)_{L,R}$ and $g_A \in \mbox{U}(1)_A$.
The nonets of scalar and pseudoscalar meson fields, $S$ and $P$, then 
parametrize the meson field $M$ as $M = S + iP$, where
\begin{align}
S & = \frac{1}{\sqrt{2}}
\begin{pmatrix}
\frac{a_0}{\sqrt{2}} + \frac{\sigma_8}{\sqrt{6}} + \frac{\sigma_0}{\sqrt{3}} & a^+ & \kappa^+ \\
a^- & -\frac{a_0}{\sqrt{2}} + \frac{\sigma_8}{\sqrt{6}} + \frac{\sigma_0}{\sqrt{3}} & \kappa^0 \\
\kappa^- & \bar{\kappa}^0 & - \frac{2\sigma_8}{\sqrt{6}} + \frac{\sigma_0}{\sqrt{3}}
\label{scalarnonet}  
\end{pmatrix},
\\ 
P & = \frac{1}{\sqrt{2}}
\begin{pmatrix}
\frac{\pi_0}{\sqrt{2}} + \frac{\eta_8}{\sqrt{6}} + \frac{\eta_0}{\sqrt{3}} & \pi^+ & K^+ \\
\pi^- & - \frac{\pi_0}{\sqrt{2}} + \frac{\eta_8}{\sqrt{6}} + \frac{\eta_0}{\sqrt{3}} & K^0 \\
K^- & \bar{K}^0 & -\frac{2\eta_8}{\sqrt{6}} + \frac{\eta_0}{\sqrt{3}}
\end{pmatrix} \,. 
\end{align}

The LSM which we employ in this paper is described by the following Lagrangian:   
\begin{align}
\mathcal{L} &= \mbox{Tr} \left(\partial _\mu M \partial ^\mu M^\dagger \right) - V
\,, 
\end{align}
with 
\begin{align}
V &= V_0 + V_{\rm anom} + V_{\rm SB} + V_{\rm SB-anom} \,. 
\label{V-form}
\end{align}
Here $V_0$ is invariant under 
the $\mbox{SU}(3)_L \times \mbox{SU}(3)_R \times \mbox{U}(1)_A$ symmetry. 
To this $V_0$, 
we include all the possible terms with dimension not larger than four:
\begin{align}
V_0 =&\mu^2 \, \mbox{Tr}\left(M M^\dagger\right) + \lambda_1 \, \mbox{Tr}\left[\left(M M^\dagger\right)^2\right] 
+ \lambda_2 \, \left[\mbox{Tr}\left(M M^\dagger\right)\right]^2 
\,, 
\label{V0}
\end{align}
where $\mu^2$ can be either positive or negative.

$V_{\rm anom}$ reflects the U(1)$_A$ anomaly, which is invariant under
the $\mbox{SU}(3)_L \times \mbox{SU}(3)_R$ symmetry, but violates U(1)$_{A}$ symmetry.
In present analysis, we include the term with lowest dimension as
\begin{align}
V_{\rm anom} = - B \, \left[ \det \left( M \right) + \det \left( M^\dag \right) \right] \ ,
\label{B-det}
\end{align}
where $B$ is a real constant with dimension one.
We note that this term generates the mass of the $\eta'$ meson.

The explicit symmetry breaking originated from the current quark masses 
is included through the mass matrix 
$
{\mathcal M} = 
{\rm diag}\{m_u, m_d, m_s\}  
$.
It is convenient to consider ${\mathcal M}$ as a spurion field 
transforming as ${\mathcal M} \to g_A \, g_L {\mathcal M} g_R^\dag$.
Then, in an ordinary manner, the current-quark mass effect is introduced in the chiral invariant way, and included in $V_{\rm SB}$.
As included in many models, the simplest term is expressed as
\begin{align}
V_{\rm SB} = & - c \, \mbox{Tr} \left [\mathcal{M}^\dag M + {\mathcal M} M^{\dagger}\right] \,,
\label{V SB}
\end{align}
where the coefficient $c$ is
a real constant having dimension two.

In the present analysis, 
we go beyond the simplest ansatz by incorporating the 
anomaly contribution depicted in Fig.~\ref{fig:SU3-mechanism} into
$V_{\rm SB-anom}$:
\begin{align} 
V_{\rm SB-anom} = & - k c  \left[ \epsilon _{abc}  \epsilon ^{def} \mathcal{M}^a_d M^b_e M^c_f + h.c. \right]
\ ,
\label{V k}
\end{align}
where $k$ is a real constant with mass-dimension minus one.
$\epsilon_{abc}$ is totally anti-symmetric under the exchange 
of indices $a$, $b$ and $c$ with $\epsilon_{123} = 1$.
Here, the summations over the repeated indices are understood. 
This term manifestly produces the contribution as described in a schematic 
graph in Fig.~\ref{fig:SU3-mechanism}, as will be clearly seen below. 
It should also be noted that the $V_{\rm SB-anom}$ is the unique term 
associated with 
the U(1)$_A$ anomaly having the SU(3) flavor breaking by
including only one ${\mathcal M}$ with the lowest dimension.

We adjust the parameters of the potential in such a way that the scalar 
nonet $S$ has a vacuum expectation value (VEV) as
$ 
\left\langle S \right\rangle = {\rm diag}\{ \alpha_1, \alpha_2, \alpha_3\}
$, 
where $\alpha_i$ ($i=1,2,3$) are real constants.
For simplicity, we assume the isospin symmetry, 
i.e., $m_u = m_d \equiv \bar{m} \neq m_s$. 
Then, the VEVs satisfy $\alpha_1 = \alpha_2 \neq \alpha_3$. 

The stationary conditions read 
\begin{align}
0 = 
& 4 \big( \mu^2 \alpha_1 + 2\lambda_1 \alpha_1^3  + 4\lambda_2 \alpha_1^3 + 2\lambda_2 \alpha_1 \alpha_3^2
 - B \alpha_1 \alpha_3 - c\bar{m} - 2kc\bar{m} \alpha_3 - 2kcm_s \alpha_1 \big) 
\,, \label{alpha1} \\ 
0 = 
& 2 \big(\mu^2 \alpha_3 + 2\lambda_1 \alpha_3^3 + 4\lambda_2 \alpha_1^2 \alpha_3 + 2\lambda_2 \alpha_3^3
 - B\alpha_1^2 - cm_s - 4kc\bar{m} \alpha_1) 
\,. \label{alpha3} 
\end{align}
The $\alpha_1$ and $\alpha_3$ are related to 
the pion decay constant $f_\pi$ and the kaon decay constant $f_K$ as
\begin{align}
f_\pi &= 2\alpha_1 \ , \label{fpi} \\
f_K &= \alpha_1 + \alpha_3 \ .
\end{align}
The quark condensates $\left<\bar{n}n\right>\equiv \left<\bar{u}u\right>= \left<\bar{d}d\right>$ 
and $\left<\bar{s}s\right>$ are calculated as
\begin{align}
\label{nbarn_condensate}
\left<\bar{n}n\right> = \left<\frac{\partial\mathcal{L}}{\partial\bar{m}}\right> &= -2c(\alpha_1 + 2k\alpha_1\alpha_3), \\
\label{sbars_condensate}
\left<\bar{s}s\right> = \left<\frac{\partial\mathcal{L}}{\partial m_s}\right> &= -2c(\alpha_3 + 2k\alpha_1^2).
\end{align}
The masses of pion and kaon are expressed as
\begin{align}
m^2_{\pi} &= \frac{c\bar{m}}{\alpha_1} (1 + 2k\alpha_3) \ , \label{pi mass} \\
m^2_{K} &= \frac{c\bar{m} + cm_s}{\alpha_1 + \alpha_3} (1 + 2k\alpha_1).
\label{K mass}
\end{align}
From Eqs.~(\ref{fpi})-(\ref{K mass}), 
one can easily see that Gell-Mann-Oakes-Renner relations 
are satisfied even in the presence of the newly introduced $k$-term in Eq.~(\ref{V k}):
\begin{align}
f_{\pi}^2 m_{\pi}^2 &= -2\bar{m}\left<\bar{n}n\right>, \\
f_{K}^2 m_{K}^2 &= -\frac{\bar{m} + m_s}{2} (\left<\bar{n}n\right> + \left<\bar{s}s\right>).
\end{align}
This must be so because the corrections to the masses 
from the newly introduced $k$-term 
can be absorbed into redefinition of the chiral condensates 
(i.e. ambiguity of its renomalization scale), 
as manifested by the 
formulas for the masses given in Eqs.~(\ref{pi mass}) and (\ref{K mass}). 

\section{Effect of Anomaly-Induced Flavor Breaking in the Meson Mass Spectra}
\label{sec:massesl}

In this section, 
we study the meson mass spectra in the LSM with the 
anomaly-induced explicit SU(3)-flavor breaking term introduced in the previous section. 

Let us first consider the contributions to the masses of the $a_0$ and $K_0^*$ mesons 
arising through $V_{\rm SB-anom}$in Eq.~(\ref{V k}).
The squared masses for the $a_0$ and $K_0^*$ mesons (before applying the vacuum conditions in Eqs.~(\ref{alpha1}) and (\ref{alpha3})) 
are calculated as
\begin{align}
\label{ma0}
m^2_{a_0} =\ & \mu^2 + 6\lambda_1 \alpha_1^2 + 2\lambda_2(2\alpha_1^2 + \alpha_3^2) + B\alpha_3 + 2k c m_s, \\
\label{mkappa}
m^2_{K_0^*}  =\ & \mu^2 + 2\lambda_1(\alpha_1^2 + \alpha_1 \alpha_3 + \alpha_3^2) + 2\lambda_2(2\alpha_1^2 + \alpha_3^2) + B\alpha_1 + 2 k c \bar{m}.
\end{align}
We shall restrict ourselves to the case 
where $\lambda_1 > 0$ and $\lambda_2 > 0$, 
which will be consistent with a favored parameter region 
shown later (Table~\ref{tab:parameters-chisq-result}).
In the case where $B=0$ and $k=0$, $m_{a_0}^2 < m_{\kappa}^2$, 
since $\alpha_3 > \alpha_1$. 
When $B \neq 0$ with $k=0$, if $\alpha_3$ were 
much larger than $\alpha_{1}$, which leads to $f_K/f_\pi$ much bigger than the experimental value, 
$m^2_{a_0} > m^2_{K_0^*}$ could be realized, as shown in Ref.~\cite{Ishida:1999}. 
However, when we use $\alpha_1 = f_\pi/2$ and $\alpha_3 = f_K - f_\pi/2$ as inputs, 
we actually have $m_{a_0}^2 < m_{\kappa}^2$ as we will show below, 
where one should note that the size of $B$ is completely fixed 
by the mass of the $\eta^{\prime}$ meson. 
Thus, one can expect that with the $k$-terms as in Eqs.~(\ref{ma0}) and (\ref{mkappa}), 
$m^2_{a_0} > m^2_{K_0^*}$ can surely be realized, 
in accordance with the schematic graph illustrated in Fig.~\ref{fig:SU3-mechanism}.  
 
To see the $k$-term effect more explicitly, 
we perform a numerical analysis for the masses of the $a_0$ and $K_0^*$ mesons.  
In Appendix~\ref{app:masses}, we explicitly 
show the formulas for the masses of the scalar and pseudoscalar mesons, 
other than $a_0, K^*_0$ (in Eqs.(\ref{ma0}) and (\ref{mkappa})) and $\pi, K$ 
(in Eqs.(\ref{pi mass}) and (\ref{K mass})), 
which are used for the numerical analysis here.
In expressing the meson masses, 
we replace parameters $\mu^2$ and $\lambda_1$ with $\alpha_1$ and $\alpha_3$  
by using the stationary conditions in Eqs.~(\ref{alpha1}) and (\ref{alpha3}).
It is interesting to note also that  
in the formulas for the masses of the $a_0$ and $K_0^\ast$ mesons in Eqs.~(\ref{ma0}) and (\ref{mkappa}) 
as well as in the stationary conditions (\ref{alpha1}) and (\ref{alpha3}),  
$\lambda_2$ appears together with $\mu^2$ necessarily by the combination $\mu^2 + 2\lambda_2(2\alpha_1^2 + \alpha_3^2) $, 
so that $\lambda_2$ can completely be removed in those mass formulas 
when the stationary conditions are used to eliminate $\mu^2$.
As a result, 
$m_{a_0}$ and $m_{K_0^*}$ are given as functions of just six parameters:~\footnote{
One cannot separate $c$ with $\bar{m}$ and $m_s$.
}
\begin{equation}
\alpha_1 \,\ \alpha_3 \,,\ c\bar{m} \,,\ c m_s \,,\ B\,,\ k \ .
\end{equation}
For a given value of $k$, we fix the values of the five parameters using the physical values of 
$f_\pi$, $f_K$, $m_\pi$, $m_K$ and $m_{\eta'}$ shown in Table~\ref{tab:PS} as inputs, 
and calculate the values of $m_{a_0}$ and $m_{K_0^*}$.
\begin{table}[t]
\centering
\caption{Input values of the masses and decay constants for pseudoscalar mesons in unit of MeV.}
\label{tab:PS}
\begin{tabular}{ccccc}
\hline
$ f_\pi$ & $f_K$ & $m_\pi$ & $m_K$ & $m_{\eta'}$\\
\hline
$92.1$ & $109$ & $138$ & $494$ & $958$\\
\hline
\end{tabular}
\end{table} 
\begin{figure}[h]
\begin{center}
\includegraphics[bb = 0 0 360 214, 
width=7.5cm]{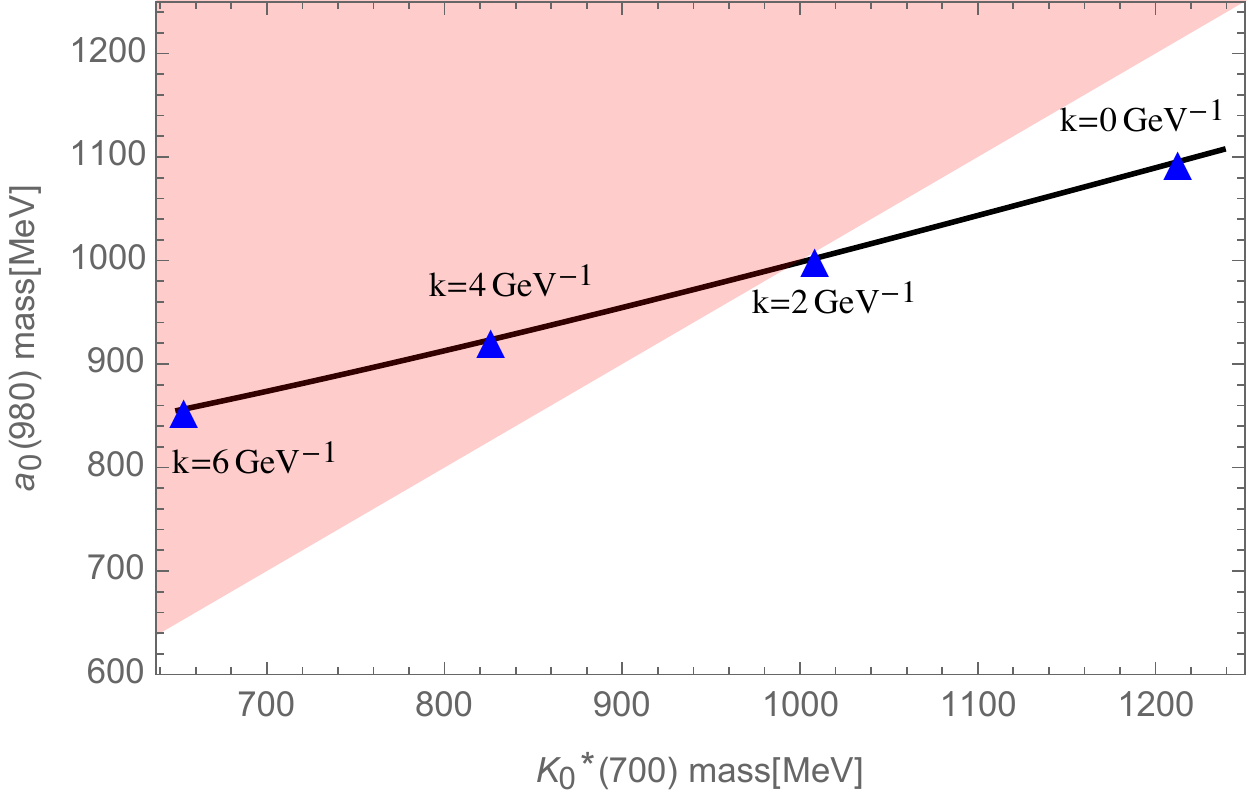}
\caption{
Correlation between $m_{a_0}$ and $m_{K_0^*}$ with values of $k$. 
In the shaded area, $m_{a_0} > m_{K_0^*}$.}
\label{fig:a0vskappa}
\end{center}
\end{figure}
Figure~\ref{fig:a0vskappa} plots $(m_{a_0}, m_{K_0^*})$ for a given $k$. 
There we find some correlation between $m_{a_0}$ and $m_{K_0^*}$ for a given $k$. 
Note that when $k = 0\ \rm{GeV^{-1}}$, the $K_0^*$-mass is larger than the $a_0$-mass, 
which does not realize the inverse hierarchy in Eq.~(\ref{inverse_hierarchy}). 
  
As $k$ develops, the mass difference $m_{a_0} -\ m_{K_0^*} $ decreases. 
Remarkably,  at $k \simeq 2.1\ \rm{GeV^{-1}}$,  
$K_0^*$ turns to be lighter than $a_0$. 
Thus, the nonzero $k$-contribution can make  
the $a_0$ and $K_0^*$ meson masses 
consistent with the inverse order as in Eq.~(\ref{inverse_hierarchy}), 
in accord with the qualitative observation depicted in Fig.~\ref{fig:SU3-mechanism}. 

Now, we show that the masses of four low-lying scalar mesons, 
$m_{a_0}$, $m_{K_0^*}$, $m_{f_0 (980)}$ and $m_{f_0 (500)}$, 
can precisely realize the desired inverse hierarchy in Eq.~(\ref{inverse_hierarchy}). 
Note that in the mass formulas for $f_0(980)$ and $f_0(500)$ given in Appendix A, 
the parameter $\lambda_2$ cannot be removed, so it is now a relevant parameter.
As was done above, 
we again use the physical values of 
$m_\pi$, $m_K$, $m_{\eta'}$, $f_\pi$ and $f_K$ listed in Table~\ref{tab:PS} as inputs, 
to get five relations among seven parameters: 
\begin{equation}
\alpha_1 \,\ \alpha_3 \,,\ c\bar{m} \,,\ c m_s \,,\ B\,,\ k \,,\ \lambda_2\ .
\end{equation}
We then fit two of these parameters to the masses of three scalar mesons 
set to the experimental values~\cite{Tanabashi:2018oca}:
\begin{align}
\label{exp_mass_scalar}
m^{\rm{(exp)}}[a_0 (980)] &= 980 \pm 20\ \rm{MeV} = m_1^{\rm{(exp)}} \pm \delta m_1^{\rm{(error)}}, \notag \\ 
m^{\rm{(exp)}}[K_0^\ast (700)] &= 824 \pm 30\ \rm{MeV} = m_2^{\rm{(exp)}} \pm \delta m_2^{\rm{(error)}}, \notag \\
m^{\rm{(exp)}}[f_0 (980)] &= 990 \pm 20\ \rm{MeV} = m_3^{\rm{(exp)}} \pm \delta m_3^{\rm{(error)}},
\end{align}
and minimize the following $\chi^2$ function:
\begin{equation}
\label{chisq-def}
\chi^2 = \sum_{i=1}^3 \frac{(m_i^{\rm{(theo.)}} - m_i^{\rm{(exp.)}})^2}{(\delta m_i^{\rm{(error)}})^2}.
\end{equation}
The best fitted values of the model parameters are listed in Table~\ref{tab:parameters-chisq-result}.
Note $\mu^2 > 0$ in Table~\ref{tab:parameters-chisq-result}, 
which implies that the spontaneous-chiral symmetry-breaking 
has been triggered essentially by a sizable cubic term coming from the 
U(1)$_A$ anomaly ($B$ term in Eq.~(\ref{B-det})), 
as was discussed in Ref.~\cite{KJKH prep}.  
Using these 
best fitted 
parameters, 
we predict the masses of the remaining scalar and pseudoscalar mesons 
as displayed in Table~\ref{tab:chisq-result}. 
From the table, we clearly see that
the inverse hierarchy for scalar mesons in Eq.~(\ref{inverse_hierarchy}) is precisely realized~\footnote{
The predicted mass of $f_0 (500)$ is slightly larger than the experimental value, 
which could however be pulled down to be 
lighter by including $\pi\pi$ rescattering effects~\cite{Bolokhov:1993}.
}.  

Using the formulas for the quark condensates $\left<\bar{n}n\right>$ and $\left<\bar{s}s\right>$ 
in Eqs.~(\ref{nbarn_condensate}) and (\ref{sbars_condensate}),
and the best fitted values of the parameters in Table~\ref{tab:parameters-chisq-result}, 
we also compute the ratio of the quark condensates to get 
$ 
\frac{\left<\bar{s}s\right>}{\left<\bar{n}n\right>} = 1.184 \pm 0.004.
$
This is consistent with the result of a recent lattice QCD analysis in Ref.~\cite{McNeile:2013}, 
$\left<\bar{s}s\right>/\left<\bar{n}n\right>|\rm{_{lat.}} = 1.08 \pm 0.16$ (at the renormalization scale $\mu=2$ GeV), 
which was not the case for the previous work~\cite{Ishida:1999}, 
as was noted in Introduction.   
Thus, the present LSM with nonzero $k$-term 
has surely improved the LSM description also for realizing 
the small enough SU(3) flavor breaking on the quark condensates~\footnote{
As to other SU(3) flavor breaking signals, 
we may also estimate the ratio of two quark masses 
using the best fitted value in Table~\ref{tab:parameters-chisq-result}, 
to get $m_s/\bar{m} = 32.3 \pm 0.1$, 
which is slightly larger than the Particle Data Group value~\cite{
Tanabashi:2018oca}:$m_s/\bar{m} = 27.3^{+0.7}_{-1.3}$. 
We expect that this will be cured by including contributions of higher order in quark masses.}
.  
\begin{table}[t]
\caption{Best fitted values of model parameters.}
\label{tab:parameters-chisq-result}
\centering
\begin{tabular}{c|c}
\hline
$\mu^2$ & (1.02 $\pm$ 2.28) $\times \rm{10}^4\ \rm{MeV}^2$\\
\hline
$\lambda_1$ & 11.8 $\pm$ 1.3 \\
\hline
$\lambda_2$ & 20.4 $\pm$ 1.3 \\
\hline
$B$ & (3.85 $\pm$ 0.03) $\times \rm{10}^3\ \rm{MeV}$\\
\hline
$c\bar{m}$ & (6.11 $\pm$ 0.06) $\times \rm{10}^5\ \rm{MeV}^3$\\
\hline
$cm_s$ & (198 $\pm$ 2) $\times \rm{10}^5\ \rm{MeV}^3$\\
\hline
$k$ & (3.40 $\pm$ 0.12) $\rm{GeV}^{-1}$\\
\hline
\end{tabular}
\end{table}
\begin{table}[t]
\caption{
Best fitted values of masses of scalar and pseudoscalar mesons (listed in the first clumn). The experimental values for
masses of the $a_0 (980)$, $K_0^\ast (700)$ and $f_0 (980)$ mesons are included in the 
fitting as inputs.
Comparisons with the experimental values regarding our predictions 
(``exp.'' in the second column)  
and the earlier work based on a similar LSM~\cite{Ishida:1999}, but 
without the $k$-term (the third column) are also presented.} 
\label{tab:chisq-result}
\centering
\begin{tabular}{c|c|c|c}
 & our model & exp. value\cite{Tanabashi:2018oca} & \cite{Ishida:1999} \\
\hline \hline
$a_0 $ & 937 $\pm$ 4 MeV & 980 $\pm$ 20 MeV & 900 - 930 MeV \\
\hline
$K_0^* (700)$ & 863 $\pm$ 10 MeV & 824 $\pm$ 30 MeV & 905$^{+65}_{-30}$ MeV \\
\hline
$f_0 (980)$ & 990 $\pm$ 18 MeV & 990 $\pm$ 20 MeV & 1030 - 1200 MeV \\
\hline \hline
$\eta$ & 552.7 $\pm$ 0.3 MeV & \begin{tabular}{c} 547.862 \\ $\pm$ 0.017 MeV \end{tabular} & input \\
\hline
$f_0 (500)$ & 672 $\pm$ 14 MeV & 400 -  550 MeV & 535 - 650 MeV \\
\hline
\end{tabular}
\end{table}
\begin{table}[t]
\centering
\caption{
Best fitted values for the squared-mass matrix elements 
in Eq.~(\ref{mass_matrix_sigma}), displayed in unit of $\rm{MeV}^2$,  
and the yielded eigenvalues in unit of MeV.
}
\label{tab:f0mix}
\begin{tabular}{c|c|c||c|c}
\hline
$m^2_{\sigma_8}$ & $m^2_{\sigma_0}$ & $|m^2_{\sigma_{08}}|$ & $m[f_0 (980)]$ & $m[f_0 (500)]$ \\
\hline
$(850 \pm 145)^2$ & $(843 \pm 210)^2$ & $(514 \pm 89)^2$ & $990 \pm 20$ & $672 \pm 14$ \\
\hline
\end{tabular}
\end{table}

In the end, we show our prediction for the 
quark contents of the $I=0$ (isosinglet) scalar mesons, 
which in the present model arises as a mixture of an $I=0$ member 
of the SU(3) flavor octet ($\sigma_8$) and a member of 
the singlet ($\sigma_0$). 
The mixing structure is read off from Eq.~(\ref{mass_matrix_sigma}) 
with the best fitted values given in Table~\ref{tab:parameters-chisq-result} used. 
The values for the mass matrix elements are listed in Table~\ref{tab:f0mix}, 
along with the mass eigenvalues corresponding to the masses 
for $f_0(980)$ and $f_0(500)$. 

The table shows that the mass matrix elements for the octet and singlet members almost degenerate
and that the mass-eigenstate scalars get highly split due to the large mixing. 
Note that this large mixing between the octet and singlet members of $I=0$ scalar mesons 
comes mainly through the ordinary type of the explicit 
SU(3) flavor breaking part $V_{SB}$ in Eq.~(\ref{V SB}): 
one can easily check it by estimating the ratio
of the second term (proportional to $k$) in Eq.~(\ref{sigma nd mass}) 
to the mixing strength with the best fitted parameters in 
Table~\ref{tab:parameters-chisq-result} 
as $\left\vert \frac{2 \sqrt{2} }{3} k c \left( \bar{m} - m_s \right) / m_{\sigma_8\sigma_0}^2 \right\vert \sim 0.2$.
Hence we expect that this mixing structure should be like so-called ``ideal mixing''. 
Indeed, when we define the mixing angle $\theta_\sigma$ by,
\begin{align}
\begin{pmatrix}
 f_0 (980)\\
 f_0 (500)
\end{pmatrix}
=
\begin{pmatrix}
 \cos\theta_\sigma & -\sin\theta_\sigma \\
 \sin\theta_\sigma & \cos\theta_\sigma
\end{pmatrix} 
\begin{pmatrix}
 \sigma_8\\
 \sigma_0
\end{pmatrix}
\ ,
\label{mixing_angle_sigma}
\end{align}
its best fitted value is 
\begin{equation}
\theta_\sigma = 44.3^\circ \pm 2.3^\circ,
\label{theta_sigma_value}
\end{equation}
which is close to $54.7^\circ $, the size which the ideal mixing yields. 
The predicted quark contents of $f_0 (980)$ and $f_0 (500)$ are summarized 
in Table~\ref{tab:Quark_Contents}.
This 
shows that $f_0 (500)$ is dominantly composted of the $\bar{n}n$ component, 
while $f_0 (980)$ is almost made of the $\bar{s}s$.

\begin{table}[t]
\caption{
Predicted 
quark contents for $f_0 (980)$ and $f_0 (500)$. 
Here $\bar{n}n$ and $\bar{s}s$ are defined by 
the diagonal elements of the matrix field $S$ in Eq.~(\ref{scalarnonet}), 
$S_1^1$ , $S_2^2$ and $S_3^3$, as 
$\bar{n}n/\sqrt{2} = S_1^1 + S_2^2$,  
and $\bar{s}s/\sqrt{2} = S_3^3$. 
}
\label{tab:Quark_Contents}
\centering
\begin{tabular}{ccc}
\hline
state & $\bar{n}n$ [$\%$] & $\bar{s}s$ [$\%$] \\
\hline
$f_0 (980)$ & 2.5 $\pm$ 1.3 & 97.5 $\pm$ 1.3 \\
$f_0 (500)$ & 97.5 $\pm$ 1.3 & 2.5 $\pm$ 1.3 \\
\hline
\end{tabular}
\end{table}

\section{Summary}
\label{sec:discussion}

In summary, 
we have proposed a new  
mechanism to realize the inverse mass hierarchy 
of the scalar mesons lighter than 1 GeV. 
The key term we have claimed is  
a U(1) axial anomaly including the 
explicit SU(3)-flavor breaking.
The term contributes to $a_{0}(980)$ 
with the strange quark mass and to $K^{\ast}_{0}(700)$ with the up or down 
quark mass due to its flavor singlet nature, 
as illustrated in Fig.~\ref{fig:SU3-mechanism}. 
To make our proposal concrete, 
we constructed a linear sigma model (LSM) by including 
the explicit SU(3)-flavor breaking term induced by the U(1) axial anomaly.
We showed that the effect from the current mass of the strange quark makes the $a_0$ meson heavier than 
the $K_0^\ast$ meson, as qualitatively expected from the  illustration in Fig.~\ref{fig:SU3-mechanism}. 
We explicitly observed that, due to the proposed mechanism, 
the inverse hierarchy of the scalar mesons lighter than 1\,GeV  given in Eq.~(\ref{inverse_hierarchy}) 
is indeed precisely realized, 
which supports to interpret them as $q\bar{q}$ meson. 
The predictions from the  LSM 
include the quark constituents 
for the $I=0$ scalar mesons: 
the quark content of $f_0 (500)$ is dominated by the 
$n\bar{n}$ component (with $n$ being up or down quark), 
and $f_0 (980)$ is by the strange quark bilinear's, $s\bar{s}$.  

In closing, we give several comments. 
The success of LSM to realize 
the inverse hierarchy is made actually by the presence of 
$\lambda_2$, $B$, and $k$ terms, which can potentially be 
suppressed by $1/N_c$ in the large-$N_c$ QCD. 
The important role played by these terms, 
corresponding to the contributions violating the quark line rule (the Okubo-Zweig-Iizuka rule~\cite{Okubo:1963,Zweig:1964,Iizuka:1966}), 
is 
consistent with the lattice analysis in Refs.~\cite{Kunihiro:2003yj,Wakayama:2014gpa}. 
Thereby, one would expect that the full QCD analysis performed in the future 
will make the $K_0^*$-mass smaller than the one estimated based on the
quenched approximation.  

We may include 
different terms including explicit symmetry breaking originated from the current quark masses
such as $Tr[\mathcal{M}M^\dagger M M^\dagger + h.c.]$ or
$Tr[\mathcal{M}M^\dagger]Tr[MM^\dagger] + h.c.$.
We expect that inclusion of them will improve the present fitting.

It is also interesting to include the effects of mixing with four-quark states 
as done in, e.g. Refs.~\cite{Fariborz:2009cq,Mukherjee:2012xn} 
into the LSM description with the presently proposed $k$ term. 
This will be pursued in the future~\cite{KHMJ}. 

Going beyond the qualitative or perturbative description based on the LSM, 
one would include rescattering effects as some nonperturbative aspects 
for strongly coupled mesons. 
Those effect have been shown to be 
important~\cite{Pelaez:2015qba,Hyodo:2010jp,SIshida:1996,Harada:1996} for estimating 
the widths of scalar mesons. 
This interesting issue will be done elsewhere. 

The anomaly-induced explicit  flavor breaking term in 
Eq.~(\ref{V k}) as well as the ordinary anomaly form of the KMT type 
could give some significant contributions to help understanding 
the nontrivial vacuum structure in QCD, and might also shed a new light 
on the analysis for dense/hot QCD. 
Those intriguing issues will also be explored in another publication.   

\section*{Acknowledgment}

We would like to thank Makoto Oka, Shoichi Sasaki and Makoto Takizawa for useful discussions and comments. 
The work of M.H. is supported in part by 
JSPS KAKENHI
Grant Number 16K05345. 
S.M. is supported partially by the National Science Foundation of China (NSFC) under Grant No. 11747308 and 11975108 , and the Seeds Funding of Jilin University (S.M.). 
The work of D.J.\ is partly supported by Grants-in-Aid for Scientific Research from JSPS (17K05449). 



\begin{thebibliography}{9}
\bibitem{Zweig:1964}
  G.~Zweig,
  ``An SU(3) model for strong interaction symmetry and its breaking''
  CERN-TH-412
\end{thebibliography}
%

\appendix

\section{Mass formulae}
\label{app:masses}

In this appendix, we present the formulas 
derived from the LSM for the masses of the pseudoscalar and scalar mesons.

After the stationary conditions in Eqs.~(\ref{alpha1}) and (\ref{alpha3}) are substituted, 
the masses of pion and kaon are given as in Eqs.~(\ref{pi mass}) and (\ref{K mass}). 

The masses for two isosinglet ($I=0$) pseudoscalar mesons are expressed by the 
$2 \times 2 $ matrix as 
\begin{align}
M^2_\eta =
\begin{pmatrix}
 m^2_{\eta_8} & m^2_{\eta_8 \eta_0} \\
 m^2_{\eta_8 \eta_0} & m^2_{\eta_0}
\end{pmatrix} \ , 
\end{align}
where~\footnote{
Note from Eqs.~(\ref{alpha1}) and (\ref{alpha3}) that 
\begin{align}
&\alpha_3 - \alpha_1 = 
\frac{c(m_s - \bar{m})(1 - 2k\alpha_1)}{\mu^2 + 2\lambda_1 (\alpha_1^2 + \alpha_1\alpha_3 + \alpha_3^2) + \lambda_2 (4\alpha_1^2 + 2\alpha_3^2) + B\alpha_1 + 2kc\bar{m}},
\notag 
\end{align}
so that 
$m^2_{\eta_8}$ and $m^2_{\eta_8 \eta_0}$ vanish in the chiral limit, $m_s = \bar{m} = 0$, 
as they should.
}
\begin{align}
m^2_{\eta_8}
= \ & \frac{1}{3\alpha_1 \alpha_3} \Big[2B\alpha_1 (\alpha_1 - \alpha_3)^2 + 2cm_s\alpha_1 \left( 1 + 2 k \alpha_3 \right) 
+ c\bar{m}\left\{ \alpha_3 + 2 k (2\alpha_1 - \alpha_3)^2  \right\} \Big] \ ,
\label{mass eta 8} \\
m^2_{\eta_0}
= \ &\frac{1}{3\alpha_1 \alpha_3} \Big[B\alpha_1 (\alpha_1 + 2\alpha_3)^2 + c m_s \alpha_1 \left( 1 + 8 k \alpha_3 \right) 
+ 2 c \bar{m} \left\{ \alpha_3 + 2 k \left( \alpha_1 + \alpha_3 \right)^2 \right\} \Big] \ ,
\label{mass eta 0} \\
m^2_{\eta_8 \eta_0}
= \ &\frac{\sqrt{2}}{3\alpha_1 \alpha_3} \Big[  -B\alpha_1 (\alpha_1 - \alpha_3)(\alpha_1 + 2\alpha_3) \notag \\
& \ \ \ \ \ {} - c m_s \alpha_1 \left( 1 -4 k \alpha_3 \right) 
+ c \bar{m} \left\{ \alpha_3 - 2 k \left( 2 \alpha_1 - \alpha_3 \right) \left( \alpha_1 + \alpha_3 \right) \right\} \Big]\ .
\label{mass eta nd}
\end{align}

For the mass formulas for the scalar mesons, it is convenient to have the forms before the stationary conditions are substituted.
The formulas for the masses of $a_0$ and $K_0^\ast$ are given in Eqs.~(\ref{ma0}) and (\ref{mkappa}).
The masses for two isosinglet $(I=0)$ scalar mesons are expressed by the 
$2 \times 2 $ matrix as 
\begin{align}
M^2_\sigma =
\begin{pmatrix}
 m^2_{\sigma_8} & m^2_{\sigma_8 \sigma_0} \\
 m^2_{\sigma_8 \sigma_0} & m^2_{\sigma_0}
\end{pmatrix} \ , 
\label{mass_matrix_sigma}
\end{align}
where
\begin{align}
m_{\sigma_8}^2 = \ & 
\mu^2 + \frac{B}{3} \left( 4 \alpha_1 - \alpha_3 \right) + 2 \lambda_1 \left( \alpha_1^2 + 2 \alpha_3^2 \right)
+ \frac{2}{3} \lambda_2 \left( 10 \alpha_1^2 - 8 \alpha_1 \alpha_3 + 7 \alpha_3^2 \right) \notag \\
& \ {} +  \frac{2}{3} k c \left( 4 \bar{m} - m_s \right) \ ,
\label{sigma 8 mass} \\
m_{\sigma_0}^2 = \ &
\mu^2 - \frac{2}{3} B \left( 2 \alpha_1 + \alpha_3 \right) + 2 \lambda_1 \left( 2 \alpha_1^2 + \alpha_3^2 \right) 
+ \frac{2}{3} \lambda_2 \left( 14 \alpha_1^2 + 8 \alpha_1 \alpha_3 + 5 \alpha_3^2 \right) \notag \\
& \ {} - \frac{4}{3} k c \left( 2 \bar{m} + m_s \right) \ ,
\label{sigma 0 mass} \\
m_{\sigma_8 \sigma_0}^2 = \ &
\sqrt{2} \left( \alpha_1 - \alpha_3 \right) \Big[ \frac{B}{3} + 2 \lambda_1 \left( \alpha_1 + \alpha_3 \right) 
+ \frac{4}{3} \lambda_2 \left( 2 \alpha_1 + \alpha_3 \right) \Big] + \frac{2 \sqrt{2} }{3} k c \left( \bar{m} - m_s \right) \ .
\label{sigma nd mass}
\end{align}

\end{document}